# Quantile-based hydrological modelling

Hristos Tyralis[1,2], Georgia Papacharalampous[3]

[1]Department of Water Resources and Environmental Engineering, School of Civil Engineering, National Technical University of Athens, Iroon Polytechniou 5, 157 80 Zografou, Greece

[2]Air Force Projects Authority, Hellenic Air Force, Mesogion Avenue 227–231, 15 561 Cholargos, Greece (https://orcid.org/0000-0002-8932-4997)

[3]Department of Engineering, Roma Tre University, Via V. Volterra 62, 00 146 Rome, Italy (https://orcid.org/0000-0001-5446-954X)

**Abstract**: Predictive uncertainty in hydrological modelling is quantified by using post-processing or Bayesian-based methods. The former methods are not straightforward and the latter ones are not distribution-free (i.e. assumptions on the probability distribution of the hydrological model's output are necessary). To alleviate possible limitations related to these specific attributes, in this work we propose the calibration of the hydrological model by using the quantile loss function. By following this methodological approach, one can directly simulate pre-specified quantiles of the predictive distribution of streamflow. As a proof of concept, we apply our method in the frameworks of three hydrological models to 511 river basins in contiguous US. We illustrate the predictive quantiles and show how an honest assessment of the predictive performance of the hydrological models can be made by using proper scoring rules. We believe that our method can help towards advancing the field of hydrological uncertainty.

**Keywords**: big data; catchment models; hydrological models; hydrological uncertainty; large-sample hydrology; point predictions; predictive uncertainty; quantile loss; quantile regression; uncertainty

## 1. Introduction

One purpose of hydrological models is to provide predictions of streamflow [1]. A split-sample scheme is implemented to assess the quality of predictions. The hydrological model is calibrated in the first segment of the sample, while the second segment is used for validation (testing) [2, 3]. The validation procedure should be applied by using information not available during the calibration procedure, while the model's performance can be assessed using some quantitative criteria (see [3] for guidelines on hydrological model validation). A key component of a hydrological model is its objective function, also called loss function, cost function or scoring function. The choice of the objective function depends on the purpose of modelling, which in most cases is to provide predictions close to the observations [1]. To this end, objective functions, such as the squared error or the Nash-Sutcliffe efficiency [4], are implemented.

Besides predicting the observations as closely (i.e. accurately) as possible, quantifying predictive uncertainty is also of high interest [5, 6]. Predictive uncertainty is defined here as the uncertainty of the hydrological predictions (simulations) [7] and can be quantified by a probability function. Quantification of uncertainty of hydrological predictions is closely related to the 20$^{th}$ unsolved problem in hydrology: "*How can we disentangle and reduce model structural/parameter/input uncertainty in hydrological prediction?*" [8].

Numerous methods have been proposed and implemented for quantifying predictive uncertainty by providing predictive quantiles or the full predictive distribution, including:

a. Post-processing (two-stage) techniques using stochastic or regression-based methods (including machine learning ones). These techniques model residual errors. Several relevant examples can be found in the literature; see [9, 10, 11, 12, 13, 14, 15, 16, 17, 18, 19, 20, 21, 22, 23, 24, 25, 26, 27, 28, 29, 30, 31, 32, 33, 34, 35, 36, 37] and the review by [38], including quantile regression, that is based on the quantile loss function. Further classification of the methods is possible, depending on specific attributes but this is out of scope here.

b. Monte-Carlo schemes; see [39, 40, 41].

c. Joint inference methods that are mostly based on Bayesian techniques; see [16, 42, 43].

d. Ensembles of predictions; see [44]; that they may need to be post-processed as well [9, 45, 46, 47].

e. Combinations of the above schemes [41, 48, 49].

Post-processing techniques are popular in forecasting applications, in which a deterministic (point) forecast by the hydrological model is delivered and the modeller conditions the predictand on the forecast as well as past input and output observations to obtain a predictive distribution [19]. In simulation mode, post-processing techniques are also applicable. Joint inference techniques, are popular in hydrological simulation and skip the two-stage procedure, e.g. using Bayesian methods [16]. In both cases (simulation and forecasting), the predictive distribution of the output [50, p.22] is delivered and can be used to quantify the predictive uncertainty.

Of special interest in hydrological modelling is the prediction of specific low or high quantiles of streamflow. This problem can be formed as the one in which the modeller receives a directive in the form of a specific statistical functional, i.e. a quantile of the predictive distribution [51]. In this case, it is critical that the objective function "*is consistent for it, in the sense that the expected score is minimized when following the directive*" ([51], see the definition of consistency of scoring functions in Methods Section 2).

Following the above conversation, the aim of this work is to solve a problem that we define as follows: "A hydrological modeller receives a directive to predict a quantile of the distribution of streamflow. How could she/he adapt its model to do so?". To this end, we propose the use of the quantile loss [52] as the objective function of the hydrological model at the requested (by the end user) quantile levels.

We note here that compared to the previous mainstream approaches to quantifying predictive uncertainty we differentiate in the following (while implications are discussed in detail in Discussion and Conclusions Sections 6 and 7):

a. Post-processing approaches: The quantile loss function is directly implemented by the hydrological model. On the other hand, post-processing methods model the residuals of the fitted hydrological model (that have been obtained by implementing a squared error type loss function), at a second stage (i.e. after applying the hydrological models), using a statistical or machine learning method.

b. Monte Carlo of Bayesian joint inference approaches: These methods are applied directly to the hydrological model; therefore, there is some resemblance with our approach. In this case, the differences of the two approaches (Bayesian joint inference and our proposed approach) are identical with those identified in the statistical literature regarding the differences between Bayesian statistics and quantile regression modelling. These differences are thoroughly discussed in Discussion Section 6.

Statistical modelling based on the quantile loss function is met frequently in the statistical literature and is well received by practitioners, e.g. in the optimization of linear-in-parameters models (see e.g. the book by [53]), neural networks [54], random forests (see e.g. the review by [55]) and boosting algorithms (see e.g. the review by [56]). Therefore, we believe that it will also be of practical interest to hydrologists. Calibration of a hydrological model with the quantile loss function has also been proposed by [57], [58] in the context of model selection and model structure deficiency assessment, however the value of the quantile loss function for predictive uncertainty assessment has been minimally examined, while here we frame assessments of predictive uncertainty in a formal framework.

The remainder of the paper is structured as follows. The methods, with emphasis on concepts related to the quantile loss function, are presented in Section 2. In the same Section, the suite of the three lumped GR (Génie Rural) hydrological models [59] used in the study is briefly presented. The data used to apply the hydrological models are presented in Section 3. In brief, this data originates from 511 river basins in the CONUS (contiguous US). The dataset description is followed by a summary and implementation directives for the proposed framework, which are provided in Section 4. The results of the application of the hydrological models are presented in Section 5. Lastly, the paper closes with the Discussion and Conclusions Sections 6 and 7, respectively.

## 2. Methods

Here, we present the concept of the quantile loss function whose use is proposed in the manuscript for supporting the application of a hydrological model for directly predicting the predictive quantiles of interest. An outline of the properties of this loss function is also provided, followed by a brief presentation of the three lumped GR hydrological models.

### 2.1 Quantile loss function

A real-valued random variable $\underline{x}$ might be characterized by its distribution function $F$ defined by:

$$F(x) := P(\underline{x} \leq x) \quad (1)$$

Then, $F^{-1}(a)$ is defined by:

$$F^{-1}(a) := \inf\{x: F(x) \geq a\} \quad (2)$$

$F^{-1}(a)$ is referred to as the $a^{\text{th}}$ quantile of $\underline{x}$, while inf denotes the infimum of a set of real

numbers. For instance, $F^{-1}(1/2)$ is the median or 0.50th quantile. In regression modelling, one minimizes the sum of absolute errors to estimate the median of the conditional distribution. The natural question then is "*are there analogs for regression of the other quantiles?*" [53, p.5]. The idea, elaborated by [52] is to apply the quantile loss function defined by eq. (3), instead of using the absolute error function:

$$L(r; x, a) := (r - x)(\mathbf{1}(x \leq r) - a) \qquad (3)$$

Here $\mathbf{1}(\cdot)$ denotes the indicator function, $x$ is the materialization of the variable $\underline{x}$, $a$ is the quantile level of interest, and $r$ is the respective predictive quantile. For $a = 1/2$, eq. (3) reduces to:

$$L(r; x, 1/2) = |r - x|/2 \qquad (4)$$

which is half the absolute error function. Obviously, when a sample is provided, the objective is to minimize the average score, i.e. the quantile loss function averaged over a fixed set of observations, equal to $\sum_i L(r_i; x_i, a)/n$, $i \in \{1, ..., n\}$.

The quantile loss function, defined by eq. (3), is positive and negatively oriented, i.e. the objective is to minimize it. Figure 1 illustrates the quantile loss function for $x = 0$ and varying predictive quantiles $r$s at quantile levels 0.05 and 0.95 (to understand how this loss function works). The quantile loss function is asymmetric. At the 0.05 quantile level (and, in general, when $a < 1/2$), it assigns a penalty to $r$ lower than the materialization, compared to its symmetric predictive quantile. The quantile loss function equals to 0, when $r = x$.

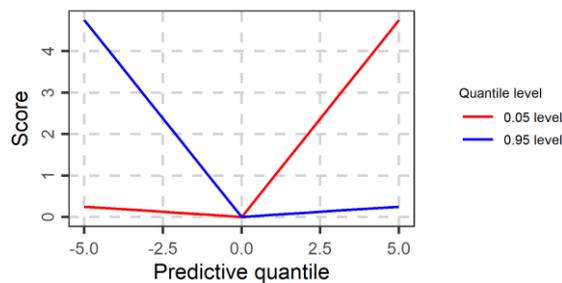

Figure 1. Illustration of the quantile score (value of the quantile loss function) at the quantile levels $a \in \{0.05, 0.95\}$ when $x = 0$ materializes and for varying predictive quantiles $r$ (see eq. (3)).

## 2.2 Theoretical properties of the quantile loss function

We note that linear-in-parameters quantile regression (or simply quantile regression when referring to non-linear models) is based on the minimization of the quantile loss function; therefore, in what follows, we refer to both the properties of quantile regression and quantile loss. A history of concepts related to quantile regression can be found in [60].

Following [51] and the intuitive explanation by [61], if a modeller "*is asked to report a certain functional of the predictive distribution, then a key requirement on the loss function is to be strictly consistent, in the sense that the expected loss or score is uniquely minimized if the directive asked for is followed*" [51]. It is proved [51] that the quantile loss function is consistent for the quantiles of the predictive distribution.

In the context of probabilistic predictions, scoring rules are used to quantify predictive performance [62]. "*A scoring rule is proper if truth telling is an optimal strategy in expectation*" [63]. Scoring rules support the modeller in being honest about the assessment of his predictions [62]. It is proved that the quantile loss function is a proper scoring rule [62].

Moving from the concepts of consistency and propriety, the quantile loss function has been the focus of intensive research, including optimization algorithms for parameter estimation in quantile regression settings [64, 65], goodness-of-fit processes [66], decomposition in reliability, resolution and uncertainty [67], and more.

## 2.3 Hydrological models

The Génie Rural GR4J, GR5J and GR6J daily lumped hydrological models were used in the study [59]. These models are widely used in hydrology, while their detailed description is out of the scope of the study. Their modular implementation in `R` programming language allows the minimization of a customized loss function.

The GR hydrological models have evolved over time from the daily GR3J model with three parameters [68] to the GR4J model with four parameters [69], and the GR5J and GR6J models with five and six parameters, respectively [70]. Other versions of the GR models are available at the monthly [71] and annual [72] time scales. The early history of the GR models can be found in [59].

The estimation of the parameters of the hydrological model is done using the Michel's [73] optimization algorithm, while the here models are calibrated and evaluated using the quantile loss function. The GR models have some interesting properties that differentiate them from one another. The additional parameter of the GR5J model with respect to the GR4J model considers more complex inter-catchment water exchanges, while the GR6J model offers improvements in the simulation of low flows [59]. Presentations of the parameters of the models can be found in [69], and [70].

## 3. Data

We applied our method to daily hydrometeorological data from 511 small- to medium-sized river basins. These river basins represent most climate types over CONUS, and their changes due to human influences are minimal. The data and detailed information about the river basins are available in the CAMELS dataset, which is fully described in [74, 75, 76, 77, 78, 79]; see Figure 2 for the geographical locations of the river basins. Daily time series of minimum and maximum temperatures and precipitation are available for each river basin [79]. The same applies for daily runoff time series. Here, we focus on the 34-year period 1980–2013. The same 34-year period and river basins have been previously examined in the hydrological post-processing study by [33], as for them the available time series records are complete. Mean daily temperatures were computed for each river basin by averaging its minimum and maximum daily temperatures, separately for each day included in the 34-year period. Lastly, a time series of daily potential evapotranspiration (PET) were estimated for each river basin by applying the Oudin's formula [80] to the previously obtained daily mean temperature time series.

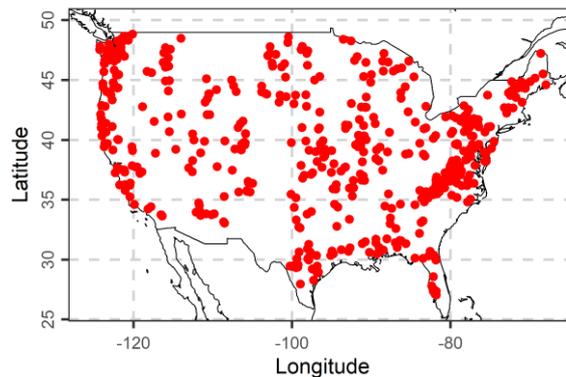

Figure 2. The 511 river basin stations over CONUS examined in the study.

## 4. Implementation and key components

Here, we present the key components and steps of the conducted large-scale application, thereby summarizing Sections 2 and 3. We applied three hydrological models, specifically the GR4J, GR5J and GR6J ones. The inputs to these models are daily precipitation and potential evapotranspiration time series, while their output is daily runoff time series. The period of interest is 1980-2013 and the dataset spans across 511 river basins in CONUS. The `airGR` R package has been used for implementing the hydrological models after appropriate adaptations. The latter facilitate hydrological model calibration based on the quantile loss function.

For an arbitrary river basin, the adopted procedures are the following:

a. Define the 2-year period 1980-1981 as the warm-up period of the hydrological models.

b. Calibrate the hydrological models in the 16-year period 1982-1997 using: (a) the quantile loss function $L(r; x, a)$ at quantile levels $a \in$ {0.025, 0.050, 0.100, 0.500, 0.900, 0.950, 0.975}; and (b) the squared error function. That equates to 8 (i.e., 7 + 1; number of loss functions) × 3 (number of hydrological models) = 24 sets of parameters at each river basin. These sets of parameters correspond to different model setups.

c. Simulate streamflow in the 16-year period 1998-2013 for each of the 24 model setups.

In what follows, the period 1998-2013 will be used for model testing purposes, while all results refer to the validation period.

## 5. Results

To facilitate the desired understanding of the outputs of the quantile-based hydrological models, in Figure 3 we present the predicted streamflow at an arbitrary river basin and for an arbitrary two-year period. We note that the models simulated streamflow in the entire 16-year period 1998-2013, and restricting the presentation of the model outputs within an arbitrary two-year period is here made for illustration purposes only. Let us now focus, for example, on the output of the GR4J model and specifically on the case that this model is calibrated at the quantile level 0.500. The simulated streamflow is close to the observed one, although the match is not perfect, since results at specific cases are subject to randomness; therefore, large-sample experiments will be presented in the following for accurate assessments. For fully perceiving this outcome, it is also sufficient to recall that the use of the $L(r; x, 0.500)$ loss function is equivalent to using the absolute error function. Instead, when the GR4J model was calibrated with the $L(r; x, 0.025)$ loss function, it simulated the streamflow quantile at level 0.025, which is lower than the observed streamflow. Similar observations and interpretations can be provided for the remaining quantile levels and hydrological models. It is also relevant to note, at this point, that the 0.025 and 0.975 quantiles contain the observed streamflow.

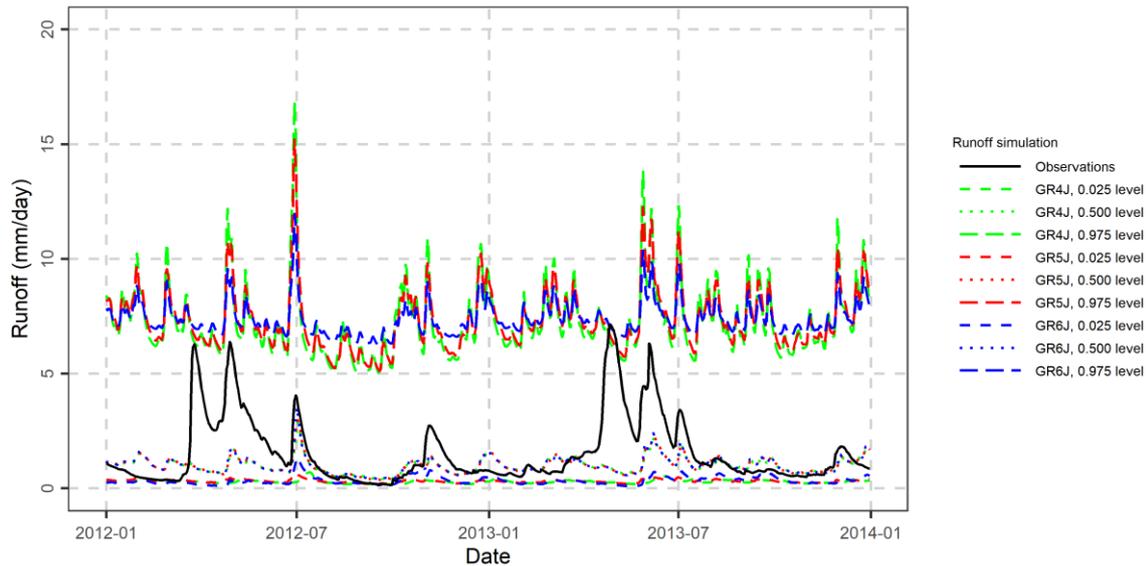

Figure 3. Illustration of observed streamflows and quantiles predicted by the GR4J, GR5J and GR6J hydrological models at the levels $a ∈ \{0.025, 0.500, 0.975\}$ for a two-year period at an arbitrary river basin.

The most frequently used objective functions in hydrological modelling are those of squared-error type. To perceive the differences between the squared error and the $L(r; x, 1/2)$ loss functions, in Figure 4 we present the average score $L(r; x, 1/2)$ in the test period at each river basin for the predictions issued by the GR4J model and specifically for the case that this model is calibrated by using the $L(r; x, 1/2)$ loss function (*x*-axis) in comparison to the squared error function (*y*-axis). It is evident that when the calibration of the hydrological model is based on the $L(r; x, 1/2)$ loss function, one receives lower average scores in the test period compared to the case in which the calibration is based on the squared error loss function; recall also that the quantile loss function is negatively oriented; therefore the lower the score the better the prediction. On the other hand, it is evident that the average scores do not differ dramatically. This is reasonable, given that both loss functions aim at simulating closely the observed streamflow.

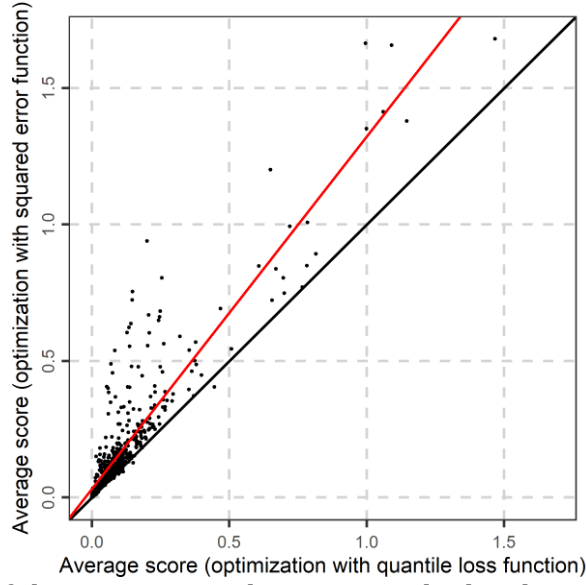

Figure 4. Scatterplot of the mean quantile scores at the level $a$ = 0.500, as computed for the testing periods and separately for each of the 511 river basins, in the case that the quantile loss function at the level $a$ = 0.500 is minimized (x-axis) and in the case that the squared error function is minimized (y-axis) for the calibration of the GR4J model.

When one performs a large-scale experiment to compare several models across a large number of river basins, illustrations such as those of Figure 3 cannot adequately support the assessment. Results should be summarized by using scaled scores. Unfortunately, quantile scores are scale-dependent; therefore, for facilitating a proper comparison a relative measure should be used. In such situations, a frequently used solution is to set a benchmark model, usually the simpler one in the experiments (which here is the GR4J model) and compare the performance of the remaining models with that of the benchmark. In particular, a relative $SCORE_{rel}$ measure can be defined by eq. (5) [81]; note the signs of the benchmark's $SCORE_{bench}$ and model's $SCORE_{model}$, reflecting that quantile loss functions are negatively oriented, therefore improvements are obtained if $SCORE_{rel}$ > 0:

$$SCORE_{rel} := (SCORE_{bench} - SCORE_{model})/SCORE_{bench} \qquad (5)$$

In Figure 5, we present the relative scores for the GR5J and GR6J models, against the average quantile score of the GR4J model. The average score refers to the loss function averaged over the test period, while it is computed separately for each set {hydrological model, calibration strategy (or targeted quantile level), river basin}. In particular, the histogram of Figure 5a presents a sample of 7 (number of quantile levels) × 511 (number of river basins) = 3 577 points, that are the relative scores of the GR5J model against the GR4J model at all the examined river basins and for the 7 examined quantile levels. The median improvement of the GR5J model against the GR4J model is 0.06%. On the other hand, the GR6J model is 1.55% worse compared to the GR4J model (see Figure 5b).

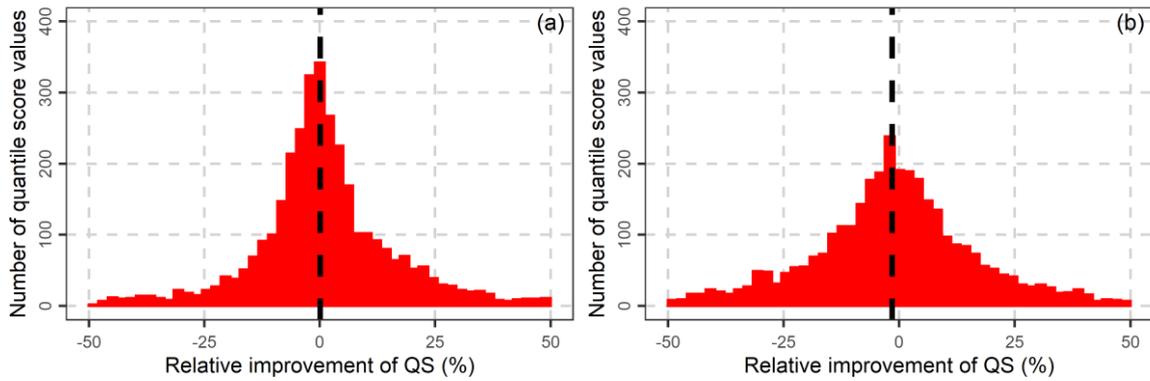

Figure 5. Histograms of the relative improvements (with red colour) and the median values of these relative improvements (with black dashed lines) against the performance of the GR4J hydrological model in terms of quantile score, as computed for all the 511 river basins and the 0.025, 0.050, 0.100, 0.500, 0.900, 0.950, 0.975 quantile levels, for the (a) GR5J and (b) GR6J hydrological models. Truncation at –50% and 50% has been applied for illustration purposes.

It is interesting to know if and how much each model improves over the performance of the benchmark at each quantile level. Figure 6 presents the median relative scores averaged over all river basins for the GR5J and GR6J models against the GR4J model. The performance of each model depends on the quantile level. For instance, the GR5J model improves over the performance of the GR4J model for quantile levels lower than 0.500, but its performance deteriorates for higher quantile levels.

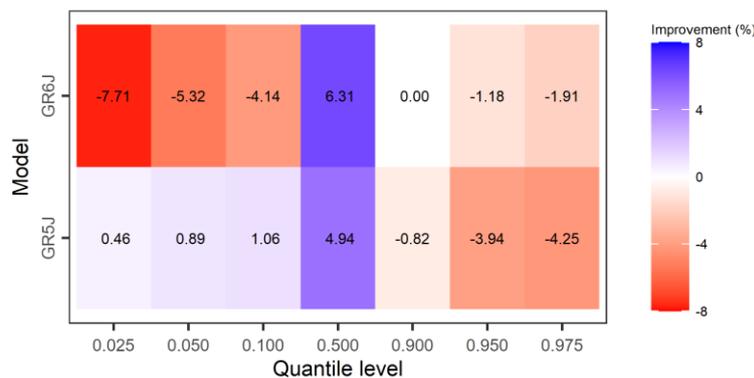

Figure 6. Heatmap of the median of the relative improvements summarizing the results for the 511 river basins for the performance of the GR5J and GR6J models against the performance of the GR4J hydrological model in terms of quantile scores.

Quantile scores are scale-dependent; therefore, intuition about the results across river basins, presented as made above (i.e. in terms of average scores), is somewhat limited. On the contrary, the coverage of the predictions issued by each model at each quantile level is a scale-independent measure. This measure counts the percentage of observations that are lower compared to the simulations at a specified quantile level. For instance, when simulating streamflow at quantile level 0.025, 2.50% of the observations will be lower than the simulated ones for a perfect model.

The median coverages of the predictions issued at all the examined river basins by the three models are provided in Figure 7. The simulations at high quantile levels seem to be better compared to those at the lower quantile levels. This could be attributed to the fact that hydrological models are less skilled to simulate low and intermittent flows. Although coverages are intuitive, we note here that they are not consistent for reporting a certain quantile of the predictive distribution of streamflow. Note that it is sufficient to recall, at this point, the definition of consistency (see Section 2.2). Therefore, they cannot support the modeller in being honest in the assessment of the hydrological models, in the sense explained in Section 2.2, and one should use proper scoring rules (e.g. the quantile score) to provide honest assessments. As an illustrative example, consider the case of obtaining predictive intervals (e.g. by combining predictive quantiles at 0.025 and 0.975 levels). One needs to have exact coverage, but also to obtain small-width intervals [82]; therefore, he has to rely on proper scoring rules.

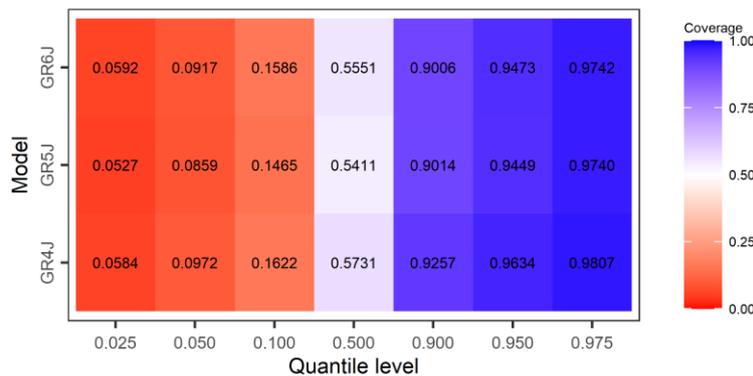

Figure 7. Heatmap of the median of the coverages of the predictions issued by the GR4J, GR5J and GR6J hydrological models at varying quantile levels summarizing the results for the 511 river basins.

## 6. Discussion

From a conceptual point of view, the differences between Bayesian approaches and our methodological approach to quantifying predicative uncertainty in hydrological modelling can be summarized by knowledge available in the field of statistics, as detailed in [83]. In particular, quantile regression (and, therefore, our methodological approach as well) is suitable in the following situations:

a. When one is interested in events at the "limits of probability".

b. When the conditional distribution does not follow a known distribution.

c. Possible presence of many outliers of the dependent variable (recall also that median regression is more robust compared to mean regression in the presence of outliers).

d. Presence of heteroscedasticity.

Obviously, streamflow attributes can be placed in the above situations; therefore, quantile-based hydrological modelling is an appropriate choice. Furthermore, compared to Bayesian approaches, quantile-based hydrological modelling is faster, as it practically constitutes an optimization problem.

On the other hand, some drawbacks of our methodological approach are the following:

a. Estimating the parameters of the model is harder compared to Gaussian regression.

b. Inference on the parameters (e.g. the computation of confidence intervals) is complicated.

c. Possible presence of quantile crossing, i.e. estimated quantiles at higher levels might be lower than respective quantiles at lower levels.

d. The full conditional distribution is not available, although the computation of multiple quantiles can substitute predictive distributions [61]. In this case, a drawback of the method is that it requires the estimation of a high number of set of model parameters (one set for each quantile).

Compared to adopting post-processing approaches, changing directly the objective function is more straightforward; therefore, the predictions issued by the proposed method could be more interpretable. In particular, post-processing approaches resemble boosting algorithms with the difference that the base models of the former (i.e. the hydrological model and the post-processors) might be strong (i.e. they might provide good predictions), albeit they do not implement the loss function of interest. Given that boosting algorithms are designed to boost performance, it might be possible that post-processing will yield better or equivalent performance in practice. This could be investigated in the future by conducting independent large experiments (see e.g. the discussions in [84, 85, 86, 87] on the value of big data experiments).

We also note that hydrological modellers are usually interested in low or high quantiles of streamflow. Although, the model's structure can be customized to target such quantities, it is frequent to combine the model with a loss function suitable for modelling means (e.g. the squared error function or the Nash-Sutcliffe efficiency [88, 89]). In such cases, one should not expect to obtain reliable estimates of quantiles of the predictive distribution, but rather customized estimates of the mean flow at low or high flow conditions.

Reporting some point predictions according to some vague request is a common practice that should be avoided, while furthermore it is not meaningful to evaluate the predictions using a set of scoring functions [51]. One should disclose the scoring function to the modeller or request a specific functional of the predictive distribution [51]. For instance, if the modeller receives a directive to report the mean functional, she/he should implement the squared error loss function which is consistent for this functional. Obviously, results could be reported for alternative scores, but they will be irrelevant for the task. Therefore, the comparison between the absolute error function and the squared error loss functions reported in Figure 4 is done for illustration purposes and is not intended to prove that one function is better with respect to the other.

Regarding the relative performance of the hydrological models at different quantile levels, more complex models provided higher performances for the 0.50 quantile although at other quantile levels a deterioration is observed with the exception of GR5J model for quantile level lower than 0.50. The reason for the performances of the models is unclear; and could be investigated in a following study. However, it should be noted that GR hydrological models were designed with a focus on the mean functional; with more complex models delivering better predictions, however these properties do not necessarily transfer to other quantile levels.

The quantile loss function can substitute the squared error function (or its equivalents) in hydrological models tailored to deliver forecasts; therefore, one can obtain directly quantile forecasts by a single model or post-process quantile simulations in the data assimilation procedure. Finally, similar methodological themes to those proposed in this work, including several ones for issuing point [90] and probabilistic [91] predictions of hydrological signatures, could be provided by exclusively using hydrological models, instead of relying on data driven-ones. Other similar themes for improved quantile-based predictions are those combining multiple hydrological models and more [30].

## 7. Conclusions

In this paper, we proposed a new methodological approach to quantifying predictive uncertainty in hydrological modelling. The key idea is to calibrate the hydrological model by using the quantile loss function. By doing so, one can directly simulate the targeted quantiles of the predictive distribution of streamflow. Compared to post-processing techniques, our approach is straightforward and, compared to Bayesian techniques, it is distribution-free and faster.

To demonstrate the effectiveness of the new method and its wide applicability far from computational limitations, we applied it to 511 river basins in CONUS in the wider frameworks of three hydrological models. With this large-scale application, we showed how the performances of different hydrological models can be compared with respect to simulating predictive quantiles at pre-defined levels. Considering its simplicity, large applicability and other advantages, we believe that our method opens new avenues in the field of hydrological modelling, with its possible extensions including those quantifying uncertainty when predicting streamflow at ungauged basins.

**Conflicts of interest:** The authors declare no conflict of interest.

**Author contributions:** The authors contributed equally to the work.

**Acknowledgements:** HT is sincerely grateful and appreciative to the Journal's award committee for having been awarded the "Water 2021 Best Paper Award" and for having been invited to submit an article to the Journal. This work has been submitted in response to this invitation.

## Appendix A    Used software

The computations and visualizations were conducted in R Programming Language [92] by using the following packages: `airGR` [59, 93], `data.table` [94], `devtools` [95], `gdata` [96], `knitr` [97, 98, 99], `rmarkdown` [100], `stringi` [101], `tidyverse` [102, 103].

## References

[1]    Solomatine DP, Wagener T (2011) 2.16 - Hydrological Modeling. In: Wilderer P (ed) Treatise on Water Science. Elsevier, pp 435–457. https://doi.org/10.1016/B978-0-444-53199-5.00044-0.

[2] Klemeš V (1986) Operational testing of hydrological simulation models. Hydrological Sciences Journal 31(1):13–24. https://doi.org/10.1080/02626668609491024.

[3] Biondi D, Freni G, Iacobellis V, Mascaro G, Montanari A (2012) Validation of hydrological models: Conceptual basis, methodological approaches and a proposal for a code of practice. Physics and Chemistry of the Earth, Parts A/B/C 42–44:70–76. https://doi.org/10.1016/j.pce.2011.07.037.

[4] Nash JE, Sutcliffe JV (1970) River flow forecasting through conceptual models part I — A discussion of principles. Journal of Hydrology 10(3):282–290. https://doi.org/10.1016/0022-1694(70)90255-6.

[5] Todini E (2007) Hydrological catchment modelling: Past, present and future. Hydrology and Earth System Sciences 11:468–482. https://doi.org/10.5194/hess-11-468-2007.

[6] Ju J, Dai H, Wu C, Hu BX, Ye M, Chen X, Gui D, Liu H, Zhang J (2021) Quantifying the uncertainty of the future hydrological impacts of climate change: Comparative analysis of an advanced hierarchical sensitivity in humid and semiarid basins. Journal of Hydrometeorology 22(3):601–621. https://doi.org/10.1175/JHM-D-20-0016.1.

[7] Montanari A (2011) 2.17 - Uncertainty of Hydrological Predictions. In: Wilderer P (ed) Treatise on Water Science. Elsevier, pp 459–478. https://doi.org/10.1016/B978-0-444-53199-5.00045-2.

[8] Blöschl G, Bierkens MFP, Chambel A, Cudennec C, Destouni G, Fiori A, Kirchner JW, McDonnell JJ, Savenije HHG, Sivapalan M, et al. (2019) Twenty-three Unsolved Problems in Hydrology (UPH) – a community perspective. Hydrological Sciences Journal 64(10):1141–1158. https://doi.org/10.1080/02626667.2019.1620507.

[9] Biondi D, Todini E (2018) Comparing hydrological postprocessors including ensemble predictions into full predictive probability distribution of streamflow. Water Resources Research 54(12):9860–9882. https://doi.org/10.1029/2017WR022432.

[10] Bock AR, Farmer WH, Hay LE (2018) Quantifying uncertainty in simulated streamflow and runoff from a continental-scale monthly water balance model. Advances in Water Resources 122:166–175. https://doi.org/10.1016/j.advwatres.2018.10.005.

[11] Bogner K, Pappenberger F (2011) Multiscale error analysis, correction, and predictive uncertainty estimation in a flood forecasting system. Water Resources Research 47(7):W07524. https://doi.org/10.1029/2010WR009137.

[12] Bogner K, Pappenberger F, Cloke HL (2012) Technical note: The normal quantile transformation and its application in a flood forecasting system. Hydrology and Earth System Sciences 16:1085–1094. https://doi.org/10.5194/hess-16-1085-2012.

[13] Bogner K, Liechti K, Zappa M (2016) Post-processing of stream flows in Switzerland with an emphasis on low flows and floods. Water 8(4):115. https://doi.org/10.3390/w8040115.

[14] Bogner K, Liechti K, Zappa M (2017) Technical note: Combining quantile forecasts and predictive distributions of streamflows. Hydrology and Earth System Sciences 21:5493–5502. https://doi.org/10.5194/hess-21-5493-2017.

[15] Dogulu N, López López P, Solomatine DP, Weerts AH, Shrestha DL (2015) Estimation of predictive hydrologic uncertainty using the quantile regression and UNEEC methods and their comparison on contrasting catchments. Hydrology and Earth System Sciences 19:3181–3201. https://doi.org/10.5194/hess-19-3181-2015.

[16] Evin G, Thyer M, Kavetski D, McInerney D, Kuczera G (2014) Comparison of joint versus postprocessor approaches for hydrological uncertainty estimation accounting for error autocorrelation and heteroscedasticity. Water Resources Research 50(3):2350–2375. https://doi.org/10.1002/2013WR014185.

[17] Krzysztofowicz R (1987) Markovian forecast processes. Journal of the American Statistical Association 82(397):31–37. https://doi.org/10.1080/01621459.1987.10478387.

[18] Krzysztofowicz R (1997) Transformation and normalization of variates with specified distributions. Journal of Hydrology 1997(1–4):286–292. https://doi.org/10.1016/S0022-1694(96)03276-3.

[19] Krzysztofowicz R (1999) Bayesian theory of probabilistic forecasting via deterministic hydrologic model. Water Resources Research 35(9):2739–2750. https://doi.org/10.1029/1999WR900099.

[20] Krzysztofowicz R (2001) The case for probabilistic forecasting in hydrology. Journal of Hydrology 249(1–4):2–9. https://doi.org/10.1016/S0022-1694(01)00420-6.

[21] Krzysztofowicz R (2002) Bayesian system for probabilistic river stage forecasting. Journal of Hydrology 268:16–40. https://doi.org/10.1016/S0022-1694(02)00106-3.

[22] Krzysztofowicz R, Kelly KS (2000) Hydrologic uncertainty processor for probabilistic river stage forecasting. Water Resources Research 36:3265–3277. https://doi.org/10.1029/2000WR900108.

[23] Li D, Marshall L, Liang Z, Sharma A, Zhou Y (2021a) Characterizing distributed hydrological model residual errors using a probabilistic long short-term memory network. Journal of Hydrology 603(Part A):126888. https://doi.org/10.1016/j.jhydrol.2021.126888.

[24] Li D, Marshall L, Liang Z, Sharma A, Zhou Y (2021b) Bayesian LSTM with stochastic variational inference for estimating model uncertainty in process-based hydrological models. Water Resources Research 57(9):e2021WR029772. https://doi.org/10.1029/2021WR029772.

[25] López López P, Verkade JS, Weerts AH, Solomatine DP (2014) Alternative configurations of quantile regression for estimating predictive uncertainty in water level forecasts for the upper Severn River: a comparison. Hydrology and Earth System Sciences 18:3411–3428. https://doi.org/10.5194/hess-18-3411-2014.

[26] Montanari A, Brath A (2004) A stochastic approach for assessing the uncertainty of rainfall-runoff simulations. Water Resources Research 40(1):W01106. https://doi.org/10.1029/2003WR002540.

[27] Montanari A, Grossi G (2008) Estimating the uncertainty of hydrological forecasts: A statistical approach. Water Resources Research 44(12):W00B08. https://doi.org/10.1029/2008WR006897.

[28] Papacharalampous GA, Tyralis H, Langousis A, Jayawardena AW, Sivakumar B, Mamassis N, Montanari A, Koutsoyiannis D (2019b) Probabilistic hydrological post-processing at scale: Why and how to apply machine-learning quantile regression algorithms. Water 11(10):2126. https://doi.org/10.3390/w11102126.

[29] Papacharalampous G, Koutsoyiannis D, Montanari A (2020a) Quantification of predictive uncertainty in hydrological modelling by harnessing the wisdom of the crowd: Methodology development and investigation using toy models. Advances in Water Resources 136:103471. https://doi.org/10.1016/j.advwatres.2019.103471.

[30] Papacharalampous GA, Tyralis H, Koutsoyiannis D, Montanari A (2020b) Quantification of predictive uncertainty in hydrological modelling by harnessing the wisdom of the crowd: A large-sample experiment at monthly timescale. Advances in Water Resources 136:103470. https://doi.org/10.1016/j.advwatres.2019.103470.

[31] Seo DJ, Herr HD, Schaake JC (2006) A statistical post-processor for accounting of hydrologic uncertainty in short-range ensemble streamflow prediction. Hydrology and Earth System Sciences Discussions 3:1987–2035. https://doi.org/10.5194/hessd-3-1987-2006.

[32] Sikorska-Senoner AE, Quilty JM (2021) A novel ensemble-based conceptual-data-driven approach for improved streamflow simulations. Environmental Modelling and Software 143:105094. https://doi.org/10.1016/j.envsoft.2021.105094.

[33] Tyralis H, Papacharalampous GA, Burnetas A, Langousis A (2019a) Hydrological post-processing using stacked generalization of quantile regression algorithms: Large-scale application over CONUS. Journal of Hydrology 577:123957. https://doi.org/10.1016/j.jhydrol.2019.123957.

[34] Weerts AH, Winsemius HC, Verkade JS (2011) Estimation of predictive hydrological uncertainty using quantile regression: Examples from the national flood forecasting system (England and Wales). Hydrology and Earth System Sciences 15:255–265. https://doi.org/10.5194/hess-15-255-2011.

[35] Yan J, Liao GY, Gebremichael M, Shedd R, Vallee DR (2014) Characterizing the uncertainty in river stage forecasts conditional on point forecast values. Water Resources Research 48(12):W12509. https://doi.org/10.1029/2012WR011818.

[36] Ye A, Duan Q, Yuan X, Wood EF, Schaake J (2014) Hydrologic post-processing of MOPEX streamflow simulations. Journal of Hydrology 508:147–156. https://doi.org/10.1016/j.jhydrol.2013.10.055.

[37] Zhao L, Duan Q, Schaake J, Ye A, Xia J (2011) A hydrologic post-processor for ensemble streamflow predictions. Advances in Geosciences 29:51–59. https://doi.org/10.5194/adgeo-29-51-2011.

[38] Li W, Duan Q, Miao C, Ye A, Gong W, Di Z (2017) A review on statistical postprocessing methods for hydrometeorological ensemble forecasting. Wiley Interdisciplinary Reviews: Water 4(6):e1246. https://doi.org/10.1002/wat2.1246.

[39] Althoff D, Rodrigues LN, Bazame HC (2021) Uncertainty quantification for hydrological models based on neural networks: The dropout ensemble. Stochastic Environmental Research and Risk Assessment 35(5):1051–1067. https://doi.org/10.1007/s00477-021-01980-8.

[40] Beven K, Binley A (1992) The future of distributed models: Model calibration and uncertainty prediction. Hydrological Processes 6(3):279–298. https://doi.org/10.1002/hyp.3360060305.

[41] Montanari A, Koutsoyiannis D (2012) A blueprint for process-based modeling of uncertain hydrological systems. Water Resources Research 48(9):W09555. https://doi.org/10.1029/2011WR011412.

[42] Hernández-López MR, Francés F (2017) Bayesian joint inference of hydrological and generalized error models with the enforcement of total laws. Hydrology and Earth System Sciences Discussions. https://doi.org/10.5194/hess-2017-9.

[43] Kuczera G, Kavetski D, Franks S, Thyer M (2006) Towards a Bayesian total error analysis of conceptual rainfall-runoff models: Characterising model error using storm-dependent parameters. Journal of Hydrology 331(1–2):161–177. https://doi.org/10.1016/j.jhydrol.2006.05.010.

[44] Vrugt JA, Robinson BA (2007) Treatment of uncertainty using ensemble methods: Comparison of sequential data assimilation and Bayesian model averaging. Water Resources Research 43(1):W01411. https://doi.org/10.1029/2005WR004838.

[45] Raftery AE, Gneiting T, Balabdaoui F, Polakowski M (2005) Using Bayesian model averaging to calibrate forecast ensembles. Monthly Weather Review 133(5):1155–1174. https://doi.org/10.1175/MWR2906.1.

[46] Reggiani P, Weerts AH (2008) A Bayesian approach to decision-making under uncertainty: An application to real-time forecasting in the river Rhine. Journal of Hydrology 356(1–2):56–69. https://doi.org/10.1016/j.jhydrol.2008.03.027.

[47] Todini E (2018) Paradigmatic changes required in water resources management to benefit from probabilistic forecasts. Water Security 3:9–17. https://doi.org/10.1016/j.wasec.2018.08.001.

[48] Koutsoyiannis D, Montanari A (2021) Bluecat: A local uncertainty estimator for deterministic simulations and predictions. https://doi.org/10.13140/RG.2.2.23863.65445.

[49] Sikorska AE, Montanari A, Koutsoyiannis D (2015) Estimating the uncertainty of hydrological predictions through data-driven resampling techniques. Journal of Hydrologic Engineering 20(1):A4014009. https://doi.org/10.1061/(ASCE)HE.1943-5584.0000926.

[50] Robert C (2007) The Bayesian Choice. Springer, New York, NY. https://doi.org/10.1007/0-387-71599-1.

[51] Gneiting T (2011) Making and evaluating point forecasts. Journal of the American Statistical Association 106(494):746–762. https://doi.org/10.1198/jasa.2011.r10138.

[52] Koenker RW, Bassett Jr G (1978) Regression quantiles. Econometrica 46(1):33–50. https://doi.org/10.2307/1913643.

[53] Koenker RW (2005) Quantile regression. Cambridge University Press, Cambridge, UK.

[54] Taylor JW (2000) A quantile regression neural network approach to estimating the conditional density of multiperiod returns. Journal of Forecasting 19(4):299–311. https://doi.org/10.1002/1099-131X(200007)19:4<299::AID-FOR775>3.0.CO;2-V.

[55] Tyralis H, Papacharalampous GA, Langousis A (2019b) A brief review of random forests for water scientists and practitioners and their recent history in water resources. Water 11(5):910. https://doi.org/10.3390/w11050910.

[56] Tyralis H, Papacharalampous GA (2021) Boosting algorithms in energy research: A systematic review. Neural Computing and Applications 33:14101–14117. https://doi.org/10.1007/s00521-021-05995-8.

[57] Pande S (2013a) Quantile hydrologic model selection and model structure deficiency assessment: 1. Theory. Water Resources Research 49(9):5631–5657. https://doi.org/10.1002/wrcr.20411.

[58] Pande S (2013b) Quantile hydrologic model selection and model structure deficiency assessment: 2. Applications. Water Resources Research 49(9):5658–5673. https://doi.org/10.1002/wrcr.20422.

[59] Coron L, Thirel G, Delaigue O, Perrin C, Andréassian V (2017) The suite of lumped GR hydrological models in an R package. Environmental Modelling and Software 94:166–171. https://doi.org/10.1016/j.envsoft.2017.05.002.

[60] Koenker RW (2017) Quantile regression: 40 years on. Annual Review of Economics 9(1):155–176. https://doi.org/10.1146/annurev-economics-063016-103651.

[61] Brehmer JR, Gneiting T (2021) Scoring interval forecasts: Equal-tailed, shortest, and modal interval. Bernoulli 27(3):1993–2010. https://doi.org/10.3150/20-BEJ1298.

[62] Gneiting T, Raftery AE (2007) Strictly proper scoring rules, prediction, and estimation. Journal of the American Statistical Association 102(477):359–378. https://doi.org/10.1198/016214506000001437.

[63] Brehmer JR, Gneiting T (2020) Properization: Constructing proper scoring rules via Bayes acts. Annals of the Institute of Statistical Mathematics 72:659–673. https://doi.org/10.1007/s10463-019-00705-7.

[64] Koenker RW, D'Orey V (1987) Computing regression quantiles. Journal of the Royal Statistical Society: Series C (Applied Statistics) 36(3):383–393. https://doi.org/10.2307/2347802.

[65] Koenker RW, D'Orey V (1994) A remark on algorithm AS 229: Computing dual regression quantiles and regression rank scores. Journal of the Royal Statistical Society: Series C (Applied Statistics) 43(2):410–414. https://doi.org/10.2307/2986030.

[66] Koenker RW, Machado JAF (1999) Goodness of fit and related inference processes for quantile regression. Journal of the American Statistical Association 94(448):1296–1310. https://doi.org/10.1080/01621459.1999.10473882.

[67] Bentzien S, Friederichs P (2014) Decomposition and graphical portrayal of the quantile score. Quarterly Journal of the Royal Meteorological Society 140(683):1924–1934. https://doi.org/10.1002/qj.2284.

[68] Edijatno, Nascimento NO, Yang X, Makhlouf Z, Michel C (1999) GR3J: A daily watershed model with three free parameters. Hydrological Sciences Journal 44(2):263–277. https://doi.org/10.1080/02626669909492221.

[69] Perrin C, Michel C, Andréassian V (2003) Improvement of a parsimonious model for streamflow simulation. Journal of Hydrology 279(1–4):275–289. https://doi.org/10.1016/S0022-1694(03)00225-7.

[70] Pushpalatha R, Perrin C, Le Moine N, Mathevet T, Andréassian V (2011) A downward structural sensitivity analysis of hydrological models to improve low-flow simulation. Journal of Hydrology 411(1–2):66–76. https://doi.org/10.1016/j.jhydrol.2011.09.034.

[71]     Mouelhi S, Michel C, Perrin C, Andréassian V (2006a) Stepwise development of a two-parameter monthly water balance model. Journal of Hydrology 318(1–4):200–214. https://doi.org/10.1016/j.jhydrol.2005.06.014.

[72]     Mouelhi S, Michel C, Perrin C, Andréassian V (2006b) Linking stream flow to rainfall at the annual time step: The Manabe bucket model revisited. Journal of Hydrology 328(1–2):283–296. https://doi.org/10.1016/j.jhydrol.2005.12.022.

[73]     Michel C (1991) Hydrologie appliquée aux petits bassins ruraux. Cemagref, Antony, France.

[74]     Addor N, Newman AJ, Mizukami N, Clark MP (2017a) Catchment attributes for large-sample studies. Boulder, CO: UCAR/NCAR. https://doi.org/10.5065/D6G73C3Q.

[75]     Addor N, Newman AJ, Mizukami N, Clark MP (2017b) The CAMELS data set: Catchment attributes and meteorology for large-sample studies. Hydrology and Earth System Sciences 21:5293–5313. https://doi.org/10.5194/hess-21-5293-2017.

[76]     Newman AJ, Sampson K, Clark MP, Bock A, Viger RJ, Blodgett D (2014) A large-sample watershed-scale hydrometeorological dataset for the contiguous USA. Boulder, CO: UCAR/NCAR. https://doi.org/10.5065/D6MW2F4D.

[77]     Newman AJ, Clark MP, Sampson K, Wood A, Hay LE, Bock A, Viger RJ, Blodgett D, Brekke L, Arnold JR, Hopson T, Duan Q (2015) Development of a large-sample watershed-scale hydrometeorological data set for the contiguous USA: data set characteristics and assessment of regional variability in hydrologic model performance. Hydrology and Earth System Sciences 19:209–223. https://doi.org/10.5194/hess-19-209-2015.

[78]     Newman AJ, Mizukami N, Clark MP, Wood AW, Nijssen B, Nearing G (2017) Benchmarking of a physically based hydrologic model. Journal of Hydrometeorology 18:2215–2225. https://doi.org/10.1175/JHM-D-16-0284.1.

[79]     Thornton PE, Thornton MM, Mayer BW, Wilhelmi N, Wei Y, Devarakonda R, Cook RB (2014) Daymet: Daily surface weather data on a 1-km grid for North America, version 2. ORNL DAAC, Oak Ridge, Tennessee, USA. Date accessed: 2016/01/20. https://doi.org/10.3334/ORNLDAAC/1219.

[80]     Oudin L, Hervieu F, Michel C, Perrin C, Andréassian V, Anctil F, Loumagne C (2005) Which potential evapotranspiration input for a lumped rainfall–runoff model?: Part 2—Towards a simple and efficient potential evapotranspiration model for rainfall–runoff modelling. Journal of Hydrology 303(1–4):290–306. https://doi.org/10.1016/j.jhydrol.2004.08.026.

[81]     Hyndman RJ, Koehler AB (2006) Another look at measures of forecast accuracy. International Journal of Forecasting 22:679–688. https://doi.org/10.1016/j.ijforecast.2006.03.001.

[82]     Dunsmore IR (1968) A Bayesian approach to calibration. Journal of the Royal Statistical Society. Series B (Methodological) 30(2):396–405. https://doi.org/10.1111/j.2517-6161.1968.tb00740.x.

[83]     Waldmann E (2018) Quantile regression: A short story on how and why. Statistical Modelling 18(3–4):203–218. https://doi.org/10.1177/1471082X18759142.

[84]     Papacharalampous GA, Tyralis H (2020) Hydrological time series forecasting using simple combinations: Big data testing and investigations on one-year ahead river flow predictability. Journal of Hydrology 590:125205. https://doi.org/10.1016/j.jhydrol.2020.125205.

[85] Papacharalampous GA, Tyralis H, Koutsoyiannis D (2019a) Comparison of stochastic and machine learning methods for multi-step ahead forecasting of hydrological processes. Stochastic Environmental Research and Risk Assessment 33(2):481–514. https://doi.org/10.1007/s00477-018-1638-6.

[86] Papacharalampous GA, Tyralis H, Papalexiou SM, Langousis A, Khatami S, Volpi E, Grimaldi S (2021) Global-scale massive feature extraction from monthly hydroclimatic time series: Statistical characterizations, spatial patterns and hydrological similarity. Science of the Total Environment 767:144612. https://doi.org/10.1016/j.scitotenv.2020.144612.

[87] Perrin C, Michel C, Andréassian V (2001) Does a large number of parameters enhance model performance? Comparative assessment of common catchment model structures on 429 catchments. Journal of Hydrology 242(3–4):275–301. https://doi.org/10.1016/S0022-1694(00)00393-0.

[88] Khatami S, Peel MC, Peterson TJ, Western AW (2019) Equifinality and flux mapping: A new approach to model evaluation and process representation under uncertainty. Water Resources Research 55(11):8922–8941. https://doi.org/10.1029/2018WR023750.

[89] Khatami S, Peterson TJ, Peel MC, Western AW (2020) Evaluating catchment models as multiple working hypotheses: On the role of error metrics, parameter sampling, model structure, and data information content. https://doi.org/10.1002/essoar.10504066.1.

[90] Tyralis H, Papacharalampous GA, Tantanee S (2019c) How to explain and predict the shape parameter of the generalized extreme value distribution of streamflow extremes using a big dataset. Journal of Hydrology 574:628–645. https://doi.org/10.1016/j.jhydrol.2019.04.070.

[91] Tyralis H, Papacharalampous GA, Langousis A, Papalexiou SM (2021) Explanation and probabilistic prediction of hydrological signatures with statistical boosting algorithms. Remote Sensing 13(3):333. https://doi.org/10.3390/rs13030333.

[92] R Core Team (2021) R: A language and environment for statistical computing. R Foundation for Statistical Computing, Vienna, Austria. https://www.R-project.org/.

[93] Coron L, Delaigue O, Thirel G, Dorchies D, Perrin C, Michel C (2021) airGR: Suite of GR hydrological models for precipitation-runoff modelling. R package version 1.6.12. https://CRAN.R-project.org/package=airGR.

[94] Dowle M, Srinivasan A (2021) data.table: Extension of 'data.frame'. R package version 1.14.2. https://CRAN.R-project.org/package=data.table.

[95] Wickham H, Hester J, Chang W (2021) devtools: Tools to make developing R packages easier. R package version 2.4.2. https://CRAN.R-project.org/package=devtools.

[96] Warnes GR, Bolker B, Gorjanc G, Grothendieck G, Korosec A, Lumley T, MacQueen D, Magnusson A, Rogers J, et al. (2017) gdata: Various R programming tools for data manipulation. R package version 2.18.0. https://CRAN.R-project.org/package=gdata.

[97] Xie Y (2014) knitr: A Comprehensive Tool for Reproducible Research in R. In: Stodden V, Leisch F, Peng RD (Eds) Implementing Reproducible Computational Research. Chapman and Hall/CRC.

[98] Xie Y (2015) Dynamic Documents with R and knitr, 2nd edition. Chapman and Hall/CRC.

[99] Xie Y (2021) knitr: A general-purpose package for dynamic report generation in R. R package version 1.36. https://CRAN.R-project.org/package=knitr.

[100] Allaire JJ, Xie Y, McPherson J, Luraschi J, Ushey K, Atkins A, Wickham H, Cheng J, Chang W, Iannone R (2021) rmarkdown: Dynamic documents for R. R package version 2.11. https://CRAN.R-project.org/package=rmarkdown.

[101] Gagolewski M (2021) stringi: Character string processing facilities. R package version 1.7.5. https://CRAN.R-project.org/package=stringi.

[102] Wickham H (2021) tidyverse: Easily install and load the 'Tidyverse'. R package version 1.3.1. https://CRAN.R-project.org/package=tidyverse.

[103] Wickham H, Averick M, Bryan J, Chang W, McGowan LD, François R, Grolemund G, Hayes A, Henry L, Hester J, Kuhn M, Pedersen TL, Miller E, Bache SM, Müller K, Ooms J, Robinson D, Paige Seidel DP, Spinu V, Takahashi K, Vaughan D, Wilke C, Woo K, Yutani H (2019) Welcome to the Tidyverse. Journal of Open Source Software 4(43):1686. https://doi.org/10.21105/joss.01686.